\RequirePackage{fix-cm}
\documentclass[twocolumn]{svjour3}          
\smartqed  
\usepackage{graphicx}
\usepackage{cite}

\usepackage{amsmath, amssymb}
\newcommand{\D}{\mathrm{d}}
\newcommand{\heff}{h_{\mathrm{eff}}}
\newcommand{\keff}{k_{\mathrm{eff}}}
\begin{document}

\title{Response functions of atom gravimeters
}


\author{V D Nagornyi}


\institute{V D Nagornyi\at
              Metromatix, Inc. \\
              \email{vn2@member.ams.org}           
}

\date{Received: date / Accepted: date}

\maketitle

\begin{abstract}
Atom gravimeters are equivalent to non-multi-level corner-cube gravimeters in translating the gravity signal into the measurement result. This enables description of atom gravimeters as LTI systems. The system's impulse responses by acceleration, velocity, and displacement are found to have the shape of triangle, meander, and the Dirac comb resp.
The effects of inhomogeneous gravity field are studied for constant and linear vertical gradients and self-attraction of the instrument. For the constant gradient the effective measurement height is below the top of the trajectory at 1/6 and 7/24 of its length for the fountain and the release types of the instruments resp.
The analysis is expanded to the gravimeters implementing the Bloch oscillations at the apex of the trajectory. In filtering the vibrations these instruments are equivalent to the first-order low-pass filters, while other atom gravimeters are equivalent to the second-order low-pass filters.
\end{abstract}

\section{Introduction}
\label{intro}
In 1991 Mark Kasevich noticed that the phase shift of the atom interferometer is quadratically proportional to the free-fall time of the atoms \cite{kasevich1991, chu1998}, thus realizing the first gravimeter with cold atoms as test mass. The following years saw rapid progress in this type of instruments, which now approach in accuracy and even exceed in sensitivity the best corner-cube gravimeters \cite{deangelis2009, gouet2008, merlet2009, louchetchauvet2011}.
The atom and the corner-cube instruments measure gravity acceleration in significantly different ways, with almost non-overlapping sets of systematic effects, so bringing both types of the instruments together for comparisons is very beneficial for metrology.
Similar to corner-cube instruments, many systematic effects of atom gravimeters can be estimated only by modelling and computer simulation. To validate the corrections for the effects it's therefore important to compare results obtained with different models.
In this paper we outline atom gravimeters from the viewpoint of the theory of linear systems. In particular, we find the impulse response functions of atom gravimeters and apply them to the analysis of the effects of the inhomogeneous gravity field.
\section{Atom gravimeter as LTI system}
\label{LTI}
\begin{figure}[t]
\centering
\input{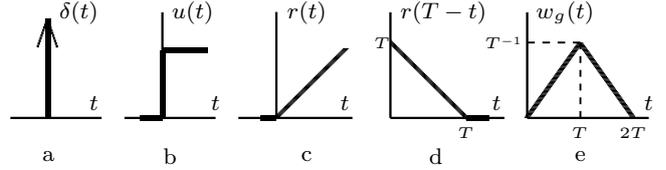}
  \caption[short title]
  {
  \quad\parbox[t]{7cm} {
The acceleration impulse response of atom gravimeter as reaction on the Dirac $\delta$-function:\\
a -- $\delta(t)$ as input signal;\\
b -- the Heaviside unit step function $u(t)=\int \delta(t) \D t$;\\
c -- the unit ramp function $r(t)= \int \int \delta(t) \D t \D t $;\\
d -- the weighting function of a double integrator as $r(t)$ turned backward w.r.t. the end of integration; \\
e -- the weighting function of atom gravimeter combined of two scaled and shifted double integrators. The process is illustrated in more details in \cite{nagornyi2011}.

    }
  }
\label{fig_AG_GENESIS}
\end{figure}

\begin{figure}[t]
\centering
\input{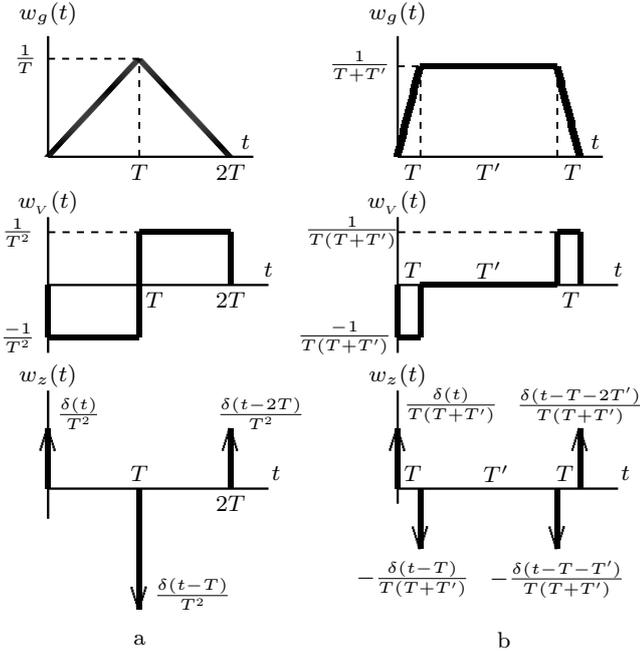}
  \caption[short title]
  {
  \quad\parbox[t]{7.5cm} {Weighting functions of atom gravimeters (top to bottom): by acceleration, by velocity, by displacement.\\
a -- 3-pulse atom gravimeter\\
b -- atom gravimeter implementing Bloch oscillations
    }
  }
\label{fig_AG_WFs}
\end{figure}

In atom gravimeters the positions of the free falling atoms are related to the phases of the light field generated by two counter-propagating lasers. In the 3-pulse gravimeters the phases $\phi_1$, $\phi_2$, $\phi_3$ of the field correspond to the positions of the atoms at the moments $t_1$, $t_2$, $t_3$ separated by the time interval $T$. The interferometer output in these gravimeters is \cite{kasevich1992, storey1994}
\begin{equation}
\label{eq_delta_phi}
\Delta \phi = k g T^2 + \phi_1 - 2\phi_2 + \phi_3,
\end{equation}
where $k$ is the wave vector\footnote{In later publications this parameter is often called the \emph{effective wave vector} $\keff$ to highlight the cumulative action of two counter-propagating beams. We use simplified notation to avoid allusions to the \emph{effective measurement height} $\heff$ considered further in this paper.}
of the light field.
The quadratic term of (\ref{eq_delta_phi}) is caused by the Doppler effect, as the accelerating atoms see the static light wave as linearly increasing frequency. The term can be cancelled by controlling the frequencies of the lasers during the free fall. The measurement of the acceleration in atom gravimeters consists in the experimental search of the frequency increase rate $\alpha$ that  compensates the Doppler shift. The acceleration is obtained from the $\alpha$ as \cite{zhou2011a}
\begin{equation}
\label{eq_g_via_alpha}
\overline g = 2\pi \alpha /k.
\end{equation}
The right $\alpha$ makes $\Delta \phi$ equal to 0 for any $T$ \cite{louchetchauvet2011}, so from (\ref{eq_delta_phi}) it also follows
\begin{equation}
\label{eq_3_phases}
\overline g = (\phi_1 - 2\phi_2 + \phi_3)/k T^2.
\end{equation}
As $z=\phi/k$, the same acceleration could also be found from the coordinates, if they were available \cite{peters1998}:
\begin{equation}
\label{eq_3_points}
\overline g = (z_3 - 2z_2 + z_1)/ T^2.
\end{equation}
Substituting the distances traveled by the atoms from the coordinate $z_1$: $S_1=z_2-z_1$, $S_2=z_3-z_1$ and the corresponding time intervals $T_1=T$, $T_2=2T$ into (\ref{eq_3_points}) reveals the equivalence of the 3-pulse atom gravimeters with the 3-level measurement schemas (fig.\ref{fig_AG_SCHEMA}) used in some corner-cube instruments, e.g. \cite{zhang2007}:
\begin{equation}
\label{eq_3_levels}
\overline g = \left(\frac{S_2}{T_2}-\frac{S_1}{T_1} \right) \frac{2}{T_2 - T_1}.
\end{equation}
Let the acceleration of atoms during the free fall change like $g(t)$. The distance relates to the acceleration via double integration, so (\ref{eq_3_levels}) can be put as
\begin{equation}
\label{eq_g_via_integrals}
\overline g = \left(\frac{\int_0^{2T} \int_0^t g(\tau) \D \tau \D t} {2T}-\frac{\int_0^{T} \int_0^t g(\tau) \D \tau \D t}{T} \right) \frac{2}{T}.
\end{equation}
The above formula relates the acceleration of atoms $g(t)$ to the measured acceleration $\overline g$ \emph{linearly}, i.e.  linear combination of partial accelerations
\begin{equation}
\label{eq_g_partials}
g(t) = a g_1(t) + b g_2(t),
\end{equation}
translates into
\begin{equation}
\label{eq_g_partials_meas}
\overline g = a \overline g_1 + b \overline g_2,
\end{equation}
where $\overline g_1$ and $\overline g_2$ are the results of independent measurements of $g_1(t)$ and $g_2(t)$.
This linearity plus the time invariance of (\ref{eq_3_levels}) (meaning that $T_1$ and $T_2$ are pre-defined for a drop) enables treatment of atom gravimeter as linear time-invariant (LTI) system . In such a system \cite{siebert1985} the input and the output signals are connected by the \emph{convolution} operation:
\begin{equation}
\label{eq_convolution}
\overline g(t) = \int_{-\infty}^{+\infty} g(\tau) h(t - \tau) \D \tau,
\end{equation}
where $h(t)$ is the \emph{impulse response function}  of the system. Applied to absolute gravimeters, the convolution enables presenting the measurement result in the form
\begin{equation}
\label{eq_g_via_wf}
\overline g = \int_0^{2T} g(t) w_g(t) \D t,
\end{equation}
i.e. as the acceleration of the atoms weighted over the measurement interval. Here we have $\overline g = \overline g(2T)$ and $w_g(t) = h(2T-t)$, as the measured gravity is attributed to the end of the measurement interval. Therefore, the gravimeter's weighting function is its impulse response function turned backwards with respect to the end of the measurement interval. The impulse response can be found as reaction of the system on the Dirac delta function $\delta (t)$ \cite{siebert1985}. As (\ref{eq_g_via_integrals}) has two double integrators, we first find the reaction of a double integrator on $\delta (t)$.  The first integral of $\delta (t)$ is the Heaviside step function $u(t)$, the second integral is the unit ramp function $r(x)$ (fig.\ref{fig_AG_GENESIS}):
\begin{equation}
\label{eq_unit_ramp}
r(t)=
\left\{
\begin{matrix}
0 & \;\; t < 0,
\\
t & \;\; t \ge 0 .
\end{matrix}
\right.
\end{equation}
The equivalence of the double integration to the convolution with the unit ramp (\ref{eq_unit_ramp}) follows from the integration by parts rule
$
\int_{0}^{T} u \, \D v = \left [uv  \right ]_0^T-\int_{0}^{T} v \, \D u
$
:
\begin{align}
&\int_{0}^{T} \underbrace{\int_{0}^{\tau}g(\tau) \,\D\tau}_u \underbrace{\D t}_{\D v} =
\\
& \left [ t\int_{0}^{t}g(\tau) \,\D \tau \right ]_0^T - \int_{0}^{T}t \, g(t) \,\D t
= \int_{0}^{T}g(t)(T-t) \, \D t. \nonumber
\end{align}
Two ramp functions combined according to the formula (\ref{eq_g_via_integrals}) yield the the gravimeter's weighting function by acceleration $w_g(t)$:
\begin{eqnarray}
\label{eq_w_g_via_ramp}
w_g(t) &= \left(\frac{r(2T-t)} {2T}-\frac{r(T-t)}{T} \right) \frac{2}{T}
\nonumber \\
&= \left\{\begin{matrix}
\frac{t}{T^2} & \;\; 0\le t \le T,
\\
\frac{2}{T} - \frac{t}{T^2} & \;\; T \le t \le 2T .
\end{matrix}\right.
\end{eqnarray}
Some systematic effects are more convenient to analyze in terms of the atoms' velocity or displacement rather than acceleration. 
Applying again integration by parts to (\ref{eq_g_via_wf}), we get
\begin{align}
\label{eq_g_via_wfV}
\overline g  & = \int_0^{2T} g(t) w_g(t) \D t
\\
& =\left [ w_g(t)\int g(t)\D t  \right ]_0^{2T}-\int_{0}^{2T} \frac{\D w_g(t)}{\D t}\int g(t)\D t \D t.  \nonumber
\end{align}
As $w_g(0) = w_g(2T)= 0$, and the velocity $V(t) = \int g(t) \D t$, the above expression simplifies to
\begin{equation}
\label{eq_g_wfV}
\overline g = \int_{0}^{2T}V(t) w_{_V}(t)\D t\D t,
\end{equation}
where
\begin{equation}
\label{eq_g_wfV_def}
w_{_V}(t) =  - \frac{\D w_g(t)}{\D t}
\end{equation}
is the weighting function by velocity\footnote{The function similar to $w_{_V}(t)$ is known in atom interferometry as \emph{sensitivity function} \cite{cheinet2008}.}. The weighting function by displacement can be found likewise, leading to
\begin{equation}
\label{eq_wz_wv_wg}
w_z(t) = -\frac{\D}{\D t}w_{{}_V}(t) =
\frac{\D^2}{\D t^2} w_g(t).
\end{equation}
These functions are shown on the fig.\ref{fig_AG_WFs}a.
The following properties hold true
\begin{equation}
\label{eq_wz_wv_wg_square}
\int_0^{2T} \!\!\!  w_g(t) \D t = \!\! 1,
\int_0^{2T} \!\!\!  w_{_V}(t) \D t = 0,
\int_0^{2T} \!\!\!  w_z(t) \D t = 0,
\end{equation}
explaining why the constant acceleration translates unchanged into the result, while the constant velocity or displacement yield no measured gravity.
\section{Analysis of systematic effects using response functions}
If the atoms' acceleration changes during the free fall like
\begin{equation}
\label{eq_g_t}
g(t) = g_0 + \Delta g(t),
\end{equation}
the additional component caused by the term $\Delta g(t)$ can be found as
\begin{equation}
\label{eq_dg_meas}
\overline{\Delta g} = \int_0^{2T} \Delta g(t) w_g(t) \D t,
\end{equation}
where $w_g(t)$ is the weighting function of the gravimeter by acceleration.
If the disturbance is expressed in terms of velocity or displacement, the additional component can be found using the corresponding weighting functions $w_{{}_V}(t)$ or $w_z(t)$  as
\begin{equation}
\label{eq_dV_meas}
\overline{\Delta g} = T^{-2} \left(
\int_T^{2T} \Delta V(t) \D t  -
\int_0^{T} \Delta V(t) \D t
\right),
\end{equation}
\begin{align}
\label{eq_dz_meas}
\overline{\Delta g}
& = T^{-2}\int_0^{2T} \Delta z(t)
\Big(
\delta(t) - 2\delta(t-T) + \delta(t-2T)
\Big) \D t \nonumber
\\
 & = \Big(
\Delta z(0) - 2 \Delta z(T) + \Delta z(2T)
\Big)T^{-2}.
\end{align}
It's interesting to observe that the formulas (\ref{eq_dz_meas}) and (\ref{eq_3_points}) are alike, as the gravimeter's weighting function by displacement is just the Dirac comb sampling the continuously changing coordinate in three points.

If the disturbance is expanded like
\begin{equation}
\label{eq_g_t_exp}
\Delta g(t) = \sum_{n=1}^N a_n t^n,
\end{equation}
its influence on the measured gravity, according to (\ref{eq_g_via_wf}), can be found as
\begin{equation}
\label{eq_dg_via_Cn}
\overline {\Delta g} = \sum_{i=n}^N a_n C_n,
\end{equation}
where $C_n$ is the $n$-th moment of the $w_g(t)$:
\begin{equation}
\label{eq_Cn}
C_n = \int_0^{2T} t^n w_g(t) \D t = \frac{2^{n+2}-2}{(n+1)(n+2)}T^n.
\end{equation}
Similar formulas can be derived also for $w_{_V}(t)$ and $w_z(t)$.
\subsection*{\textbf{Example 1.} Constant vertical gravity gradient and the effective measurement height}
\begin{figure}[t]
\vspace{-1.3cm}
\centering
\input{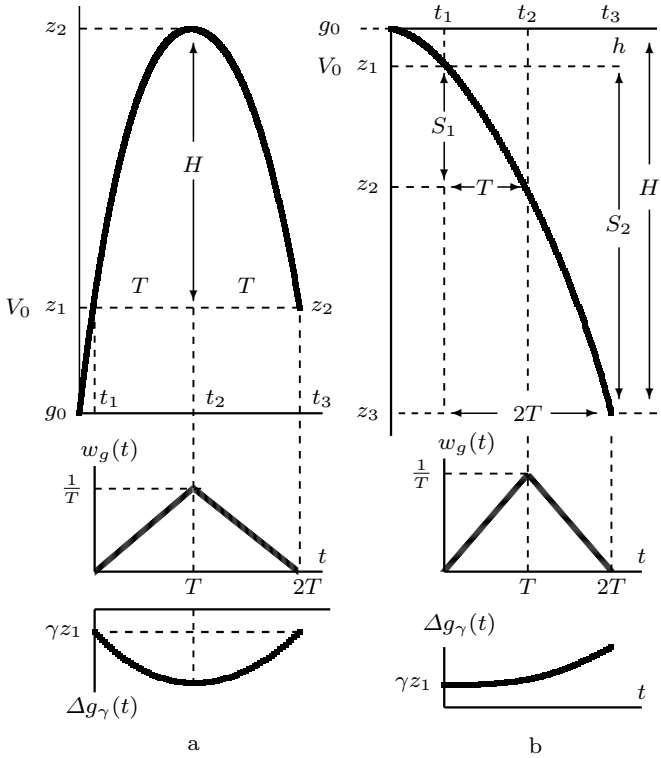}
  \caption[short title]
  {
  \quad\parbox[t]{7cm} {Trajectory model of atom gravimeters: \\
a -- fountain type, b -- release  type.\\
 Also shown are weighting functions by acceleration $w_g(t)$ and the disturbance caused by the constant gradient $\Delta g_{\gamma}(t)$.
    }
  }
\label{fig_AG_SCHEMA}
\end{figure}

We consider two realizations of atom gravimeters. In the \emph{fountain} gravimeters the measurement takes place on both upward and downward parts of the trajectory, each part has the duration of $T$ (fig.\ref{fig_AG_SCHEMA}a). In the \emph{release} gravimeters the measurement takes place only on the downward part of the trajectory during the time interval $2T$ (fig.\ref{fig_AG_SCHEMA}b).
For both types we put the coordinate origin in the atoms' resting position and co-direct $z$-axis with the atoms' initial velocity. The measurement in both cases starts at the moment $t_1$ when separation from the origin is $z_1$ and the initial velocity is $V_0$. The constant gradient changes gravity with height like
\begin{equation}
\label{eq_g_z}
g(z) = g_0 \pm \gamma \; z.
\end{equation}
From here on, the upper sign in the $\pm$ or $\mp$ symbols correspond to the release gravimeters, while the lower sign corresponds to the fountain gravimeters. In the time domain the (\ref{eq_g_z}) is
\begin{equation}
\label{eq_g_z_t}
g(t) = g_0 \pm \gamma (z_1 + V_0 t \pm g_0 t^2/2),
\end{equation}
which according to (\ref{eq_dg_via_Cn}) and (\ref{eq_Cn}) translates into the following measured gravity:
\begin{equation}
\label{eq_g_z_t}
\overline g = g_0 \pm \gamma \left (z_1 + V_0 T \pm \frac{7}{12}g_0 T^2 \right ).
\end{equation}
This agrees with the result obtained also in \cite{peters1998, rothleitner2012}.
At the \emph{effective measurement height} $\heff$ the measured gravity equals actual gravity:
\begin{equation}
\label{eq_g_z_meas}
\overline g = g_0 \pm \gamma \; \heff.
\end{equation}
Comparison of (\ref{eq_g_z_meas}) and (\ref{eq_g_z_t}) leads to the conclusion that the effective measurement height of atom gravimeters is
\begin{equation}
\label{eq_heff}
\heff = z_1 + V_0 T \pm \frac{7}{12}g_0 T^2 .
\end{equation}
%
%
For the release gravimeters, let $h$ be the length of the idle part of the trajectory, and $H$ be the total trajectory length. Obviously, $h=z_1$, $H= z_3$. As $V_0 = g_0 z_1$, $T=(t_3 - t_1)/2 $, $t_1^2 = 2z_1/g_0$, $t_3^2 = 2z_3/g_0$, we get
\begin{equation}
\label{eq_h_z0}
h_{\bf{eff}} = \frac{7}{24}(H+h) + \frac{5}{12}\sqrt{Hh}.
\end{equation}
For the fountain gravimeters, the initial velocity $V_0$ equals $g_0T$, as the atoms reach the apex in time $T$. The total trajectory height $H=g_0 T^2/2$ leads to the following location of $\heff$ above the initial position:
\begin{equation}
\label{eq_g_heff_fountain}
\heff = \frac56 H + z_1.
\end{equation}
As the gravity at the apex of the trajectory is $g_0 - \gamma(H+z_1)$, the same point is at the following distance below the apex:
\begin{equation}
\label{eq_g_sym_heff}
\heff'=\frac16 H.
\end{equation}
The gravity at the effective measurement height corresponds to the measurement with no vertical gradient correction.
\subsection*{\textbf{Example 2.} Linear vertical gravity gradient}
At some gravimetric sites the vertical gravity gradient varies significantly over the free-fall trajectory and cannot be considered a constant. The linear gradient $\gamma_1 + \gamma_2 z$ changes gravity like
\begin{equation}
\label{eq_g_z_lin_grad}
g(z) = g_0 \pm (\gamma_1 + \gamma_2 z) z.
\end{equation}
Unlike the constant gradient, the parameters $\gamma_1$ and $\gamma_2$ depend on the coordinate system used to analyze the correction. Expressing (\ref{eq_g_z_lin_grad}) in terms of time gives us
\begin{equation}
\label{g_g2_expanded}
g(t)=g_0 + A_0+A_1t + A_2t^2 + A_3t^3 + A_4t^4,
\end{equation}
where
\begin{align*}
\label{g_g2_Ai}
A_0 & = \pm (\gamma_1 z_1 + \gamma_2 z_1^2) , \; A_1 = \pm(\gamma_1 V_0 + 2\gamma_2 V_0 z_1) \\ \nonumber
A_2 & = \gamma_1 g_0/2 + \gamma_2 g_0 z_1 \pm \gamma_2 V_0^2 \\ \nonumber
A_3 & = \gamma_2 g_0 V_0 t^3 , \;\;\;
A_4  = \pm \gamma_2 g_0^2/4. \\ \nonumber
\end{align*}
Applying the moments (\ref{eq_Cn}) to (\ref{g_g2_expanded}) we obtain the effect of the linear gravity gradient as
\begin{equation}
\label{g_g2_effect}
\overline{\Delta g} = A_0 + A_1 T + \frac76 A_2 T ^2 + \frac32 A_3 T^3 + \frac{31}{15} A_4 T^4.
\end{equation}
\subsection*{\textbf{Example 3.} Self-attraction of atom gravimeter}
\begin{figure}[t]
\centering
\small
\includegraphics[height=75mm]{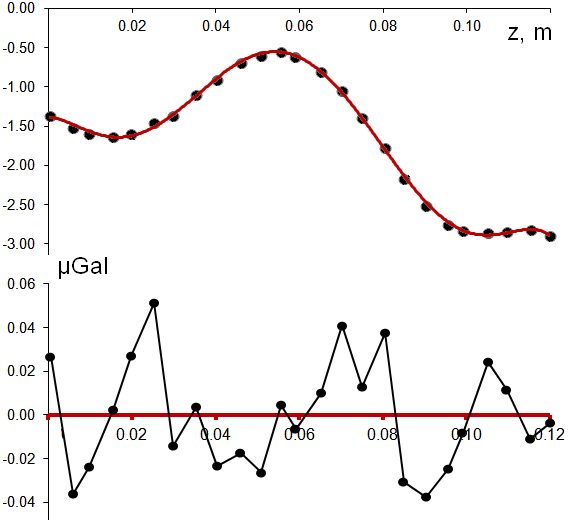}
  \caption[short title]
  {
  Disturbance caused by the self-attraction \cite{dagostino2011} and its approximation by the 6-th degree polynomial (upper plot), and the residuals of the approximation (lower plot).
  }
\label{fig_SA}
\end{figure}

The self-attraction of absolute gravimeter can produce a complex-shaped disturbance, for which the approximations by low-degree polynomials, like in the previous examples, are not sufficient. The figure \ref{fig_SA} shows the self-attraction of the atom gravimeter \cite{dagostino2011} along the atoms' trajectory and its approximation by the 6-th  degree polynomial.
To evaluate the effect of this disturbance we used moments $C_n$
up to the $n$=12. The calculated effect for the inital velocity $V_0$=0.25 ms$^{-1}$ was $-$1.27 $\mu$Gal, which agrees with the result obtained in \cite{dagostino2011} by another method. The uncertainty introduced by the approximation can be estimated with formula (\ref{eq_g_via_wf}) by treating the residuals $\epsilon (t)$ (fig.\ref{fig_SA}) as disturbance:
\begin{equation}
\label{eq_resid_est}
\int_0^{2T} \!\!\!\!\! \epsilon (t) w_g(t) \D t
\le \max |\epsilon (t)| \!\!\! \int_0^{2T} \!\!\!\!\! w_g(t) \D t
= \max |\epsilon (t)|.
\end{equation}
The result holds true for any type of absolute gravimeter: \emph{
the uncertainty of the correction caused by the approximation of the gravity disturbance does not exceed the maximum error of the approximation.
} In our case (fig.\ref{fig_SA}) the approximation led to the error of less than 0.05 $\mu$Gal.
For atom gravimeters the approximation can be performed so that the correction uncertainty would not exceed  $\max|\epsilon (t)|/4$. The detailed analysis of this issue is out of the scope of the present publication and will be presented elsewhere.
\section{Response functions of atom gravimeter implementing Bloch oscillations}
In this type of gravimeters \cite{charriere2012} the atoms are thrown up vertically and travel to the apex of the trajectory during the time $T$, where they hover for the time $T'$ tossed by the Bloch oscillations, and then drop back to the original position. The formula (\ref{eq_3_phases}) for this instrument becomes \cite{charriere2012}
\begin{equation}
\label{eq_4 phases}
\overline g = (\phi_1 - \phi_2 - \phi_3 + \phi_4)/[k T(T+T')].
\end{equation}
As per section \ref{LTI}, this gravimeter can be modeled by three double integrators with integration times of $T$, $T+T'$, and $2T+T'$. Impulse response functions of this gravimeter are shown on the fig.\ref{fig_AG_WFs}b. As $T\ll T'$, the gravimeter performs near-uniform averaging of the atoms' acceleration. The instrument described in \cite{charriere2012} has the trajectory length of only 0.8 mm, so the corrections for the constant and the linear gravity gradients are minuscule, and the correction for the self-attraction can be taken as the disturbance in any point of the trajectory taken with the opposite sign.

\begin{figure}[t]
\centering
\small
\includegraphics[height=45mm]{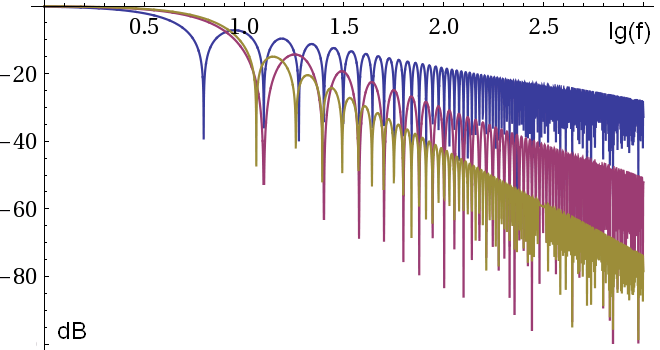}
  \caption[short title]
  {
  Amplitude of the frequency response function of the atom gravimeter with Bloch oscillations (upper plot) compared to the regular 3-pulse atom gravimeter (middle plot). Lower plot: corner-cube gravimeter with multiple levels. Total measurement interval: 0.16 s.
  }
\label{fig_FR}
\end{figure}

The gravimeter's frequency response is Fourier transform of its impulse response function \cite{svitlov2012}. Due to its almost-uniform impulse response, the atom gravimeter with Bloch oscillations is equivalent to the first-order low-pass filter. By comparison (fig.\ref{fig_FR}), the 3-pulse atom gravimeter is equivalent to the second order, while the corner-cube gravimeter is equivalent to the third-order low-pass filter \cite{svitlov2012} (fig.\ref{fig_FR}). These characteristics follow from the gravimeters' logic in translating the input gravity to the measurement result and do not include any additional vibration shielding that instruments may possess.
\vspace{-0.5cm}
\section{Conclusions}
We analyzed atom gravimeters as LTI systems, found their impulse response functions and applied them to evaluation of certain disturbances. The following conclusions sum up the analysis.
\begin{enumerate}
\item Atom gravimeters are equivalent to non-multi-level corner-cube gravimeters in translating the gravity signal into the measurement result.
\item Weighting function of atom gravimeter by acceleration can be determined as impulse response of the LTI system consisting of several double integrators. The acceleration, velocity, and displacement weighting functions are successive derivatives changing sign on every succession.
\item The effect of a disturbance on atom gravimeter can be found by replacing the time powers in the disturbance expansion with corresponding moments of the weighting function.
\item The effective measurement height of atom gravimeters is located below the apex of the trajectory on 1/6 of its total length for the fountain, and on about 7/24 ditto for the release type gravimeters.
\item Error in the analysis of a disturbance arising from the disturbance approximation does not exceed the maximum error of the approximation.
\item With respect to the vibration disturbances the atom gravimeter with Bloch oscillations is equivalent to the first-order low-pass filter, while the 3-pulse atom gravimeter is equivalent to the second order low-pass filter.
\end {enumerate}
%
%
%
%
%

%
\end{document}